\documentclass[aps,twocolumn,superscriptaddress]{revtex4-2}

\usepackage{amsmath}
\usepackage{mathtools} 
\usepackage{amssymb}
\usepackage{amsthm}
\usepackage{bm,upgreek} 
\usepackage{upgreek} 
\usepackage{xcolor}
\usepackage{graphicx}
\usepackage{epstopdf}
\usepackage[colorlinks,linkcolor=blue,anchorcolor=blue,citecolor=blue,urlcolor=blue]{hyperref}
\usepackage{tikz}
\usepackage[compat=1.1.0]{tikz-feynman}

\newcommand{\bea}{\begin{eqnarray}}
\newcommand{\eea}{\end{eqnarray}}

\newcommand{\md}{\mathrm{d}}

\newcommand{\w}{\omega}

\newcommand{\W}{\Omega}

\begin{document}

\title{Effect of anisotropic impurity scattering in d-wave superconductors}
\author{Ze-Long Wang}
\affiliation{National Laboratory of Solid State Microstructures $\&$ School of Physics, Nanjing University, Nanjing 210093, China}
\author{Rui-Ying Mao}
\affiliation{National Laboratory of Solid State Microstructures $\&$ School of Physics, Nanjing University, Nanjing 210093, China}
\author{Da Wang}\email{dawang@nju.edu.cn}
\affiliation{National Laboratory of Solid State Microstructures $\&$ School of Physics, Nanjing University, Nanjing 210093, China}
\affiliation{Collaborative Innovation Center of Advanced Microstructures, Nanjing University, Nanjing 210093, China}
\author{Qiang-Hua Wang}\email{qhwang@nju.edu.cn}
\affiliation{National Laboratory of Solid State Microstructures $\&$ School of Physics, Nanjing University, Nanjing 210093, China}
\affiliation{Collaborative Innovation Center of Advanced Microstructures, Nanjing University, Nanjing 210093, China}

\begin{abstract}
In d$_{x^2-y^2}$-wave superconductors, the effect of s-wave point disorder has been extensively studied in the literature.
In this work, we study the anisotropic disorder with a the form of $V_{\0k\0k'}^{\rm imp}=V_if_\0kf_{\0k'}$ with $f_\0k=\cos(2\theta)$ (with $\theta$ the azimuthal angle of $\0k$), as proposed to be caused by apical oxygen vacancies in overdoped La-based cuprate films, under the Born approximation.
The disorder self-energy and d-wave pairing affect each other and have to be solved simultaneously self-consistently. We find the self-energy is reduced at low frequencies and thus weakens the pair-breaking effect.
This frequency-dependence vanishes in the dirty limit for which the disorder is well described by a scattering rate $\Gamma_\0k=\Gamma_if_\0k^2$.
One consequence of the disorder effect is the gap-to-$T_c$ ratio $2\Delta(0)/T_c$ is greatly enhanced by the d-wave disorder, much larger than the s-wave disorder and the clean BCS value $4.28$.
At last, we generalize the d-wave scattering rate to a general form $\Gamma_\theta=\Gamma_\alpha|\theta-\theta_0|^\alpha$ around each nodal direction $\theta_0$.
We find the density of states $\rho(\w)-\rho(0)\propto|\w|$ ($\w^2$) for all $\alpha\ge1$ ($\alpha<1$) in the limit of $\w\to0$.
As a result, the superfluid density $\rho_s$ exhibits two and only two possible scaling behaviors: $\rho_s(0)-\rho_s(T)\propto T$ ($T^2$) for $\alpha\ge1$ ($\alpha<1$) in the low temperature limit.
\end{abstract}

\maketitle

Disorders are always inevitable in real-world superconductors and have to be taken into account carefully to interpret experiments.
In s-wave superconductors, non-magnetic impurities are found to have no effect on the transition temperature $T_c$ \cite{Anderson1959} but can reduce the superfluid density $\rho_s$, whose zero temperature value $\rho_s(0)\propto T_c$ in the dirty limit \cite{AG1958,AG1959}.
On the other hand, magnetic impurities induce in-gap bound states \cite{Yu1965} and thus causes pair-breaking. As a result, both $T_c$ and $\rho_s$ are reduced \cite{AG1961}.
For unconventional superconductors, due to the sign change of the pairing function, both non-magnetic and magnetic impurities can cause pair-breaking and reduce $T_c$ and $\rho_s$ simultaneously.

The effect of s-wave point disorder in d$_{x^2-y^2}$-wave (abbreviated as d-wave in the following) superconductors with the pairing function $\Delta_\theta=\Delta\cos(2\theta)$, where $\theta$ is the azimuthal angle relative to the antinodal direction, has been extensively investigated in the studies of high temperature cuprate superconductors \cite{RMP2009,XiangBook}.
The most significant feature of the s-wave point disorder is the low energy density of states (DOS) $\rho(\w)$ depends on $\w$ quadratically rather than linearly as in clean d-wave superconductors.
This power law scaling behavior has many consequences in experiments, such as the quadratic temperature dependence of the specific heat $C/T$, superfluid density $\rho_s$ and penetration depth $\lambda$.
Recently, such an expected power law behavior is found to be inconsistent with the experimental observation of $\rho_s(0)-\rho_s(T)\propto T$ in overdoped La$_{2-x}$Sr$_x$CuO$_4$ films \cite{Bozovic2016}, suggesting that the samples are in the clean limit. However, the same samples appear to be in the dirty limit from the observation of the scaling law $\rho_s(0)\propto T_c$ \cite{Bozovic2016} and the Drude-like optical conductivity below $T_c$ \cite{Mahmood2019}.
In order to reconcile this paradox, we proposed a d-wave anisotropic scattering rate $\Gamma_\theta=\Gamma_d\cos^2(2\theta)$ \cite{Wang2022}, which has the same form as the ``cold spot'' model \cite{Ioffe1998} but is caused by the ubiquitous apical oxygen vacancies in overdoped La-based cuprate films \cite{Sato2000,Kim2017}.
The most important feature of this anisotropic scattering rate is that it drops to zero along the nodal directions of the pairing function.
As a result, it does not smear out the low energy quasiparticle excitations, giving rise to $\rho(\w)-\rho(0)\propto|\w|$ and thus $\rho_s(0)-\rho_s(T)\propto T$.
But it does affect the ground state property such as $\rho(0)$ and $\rho_s(0)$.
Therefore, the above paradox is naturally resolved \cite{Wang2022}.

In our previous study, we did not consider the frequency dependence of the disorder self-energy $\Sigma(\w,\theta)$ affected by the d-wave pairing but only use its normal state value $\Sigma(\w,\theta)=-i\Gamma_d \cos^2(2\theta)$ corresponding to the d-wave scattering rate $\Gamma_\theta=\Gamma_d\cos^2(2\theta)$, called scattering rate approximation, which tends to be exact in the dirty limit $\Gamma_d\gg\Delta$.
In this work, we go beyond the dirty limit by solving the self-energy $\Sigma(\w,\theta)$ and d-wave pairing $\Delta$ simultaneously.
In doing so, we justify the scattering rate approximation in the dirty limit, and unravel new effects in the general cases.
We find $-\mathrm{Im}\Sigma(\w,\theta)$ is reduced at low frequency but grows up with the disorder strength $\Gamma_d$ and finally approaches the normal state scattering rate $\Gamma_d\cos^2(2\theta)$ in the dirty limit.
We find the gap-to-$T_c$ ratio $2\Delta(0)/T_c$ is greatly enhanced by the d-wave disorder, much larger than the s-wave disorder and the BCS prediction of $4.28$ in the clean limit.
We also generalize the d-wave scattering rate to $\Gamma_\theta=-i\Gamma|\theta-\theta_0|^\alpha$ near each gap nodal direction $\theta_0$.
In the scattering rate approximation, we find the DOS $\rho(\w)-\rho(0)\propto|\w|$ ($\w^2$) for all $\alpha\ge1$ ($\alpha<1$) in the limit of $\w\to0$.
As a consequence, the scaling of the low temperature superfluid density falls into two and only two categories: $\rho_s(0)-\rho_s(T)\propto T$ ($T^2$) for $\alpha\ge1$ ($\alpha<1$) in the limit of $T\to0$.

In this work, we consider the non-magnetic impurity scattering potential
\begin{align}
V_{\0k\0k'}^{\rm imp}=V_if_\theta f_{\theta'} ,
\end{align}
where $\theta$ ($\theta'$) is the azimuthal angle of the momentum $\0k$ ($\0k'$).
The potential value $V_i=V_s$ for the s-wave point disorder with $f_{\theta}=1$, and $V_d$ for the d-wave disorder with $f_{\theta}=\cos(2\theta)$.
In fact, through partial-wave decomposition, any impurity potential $V_{\0k\0k'}$ can be decoupled into different scattering channels as $V_{\0k\0k'}=\sum_n V_n f_{n\0k} f_{n\0k'}$.
In this work, for simplicity, we focus on the pure d-wave disorder and compare it with the pure s-wave point disorder well established previously in the literature \cite{XiangBook}.
Since the d-wave disorder mainly comes from apical oxygen vacancies out of the CuO$_2$-plane, its impurity potential strength $V_d$ is expected to be quite small.
Therefore, we work in the Born limit throughout this study.

In the normal state, the disorder contributes a self-energy $\Sigma_n(\w_n,\theta)=-i \text{sgn}(\w_n) \Gamma_i f_\theta^2=\Sigma_n(\w_n) f_\theta^2$ corresponding to a scattering rate $\Gamma_\theta=\Gamma_i f_\theta^2$, where $\Gamma_i=\pi n_i\+NV_i^2$, $\w_n$ is the Matsubara frequency, and $\+N$ is the normal state DOS at the Fermi energy.
In the superconducting state, the existence of d-wave pairing $\Delta$ will affect $\Sigma(\w_n,\theta)$ in particular at low frequencies.
As a result, the frequency dependence of $\Sigma(\w_n,\theta)$ will also affect the pairing $\Delta$.
Therefore, we need to self-consistently solve $\Sigma(\w_n,\theta)$ and the pairing gap $\Delta$ simultaneously \cite{AG1961, AGDBook, RMP2009}.
For this purpose, we introduce the BCS-type pairing interaction
\begin{align} \label{eq:BCS}
V_{\0k\0k'}^{\rm pair}=V\phi_\theta\phi_{\theta'}\Theta(\W-|\w_n|)\Theta(\W-|\w_n'|) ,
\end{align}
where $\phi_\theta=\cos(2\theta)$, $\Theta(\cdot)$ is the Heaviside step function and $\W$ is the frequency cutoff.
Exactly speaking, for boson mediated pairing interaction, we should require $|\w_n-\w_n'|<\W$ as in the strong coupling Eliashberg theory \cite{Eliashberg1960}.
Here, we take Eq.~\ref{eq:BCS} as a good approximation \cite{BCS1957}.
We substitute the bare Green's function $g^{-1}=i\w_n\sigma_0-\varepsilon_\0k\sigma_3$ and full Green's function $G^{-1}=i\tilde{\w}_n\sigma_0-\tilde{\varepsilon}_\0k\sigma_3-\Delta_\0k\sigma_1$ with $i\tilde{\w}_n(\w_n,\theta)=i\w_n-\Sigma(\w_n,\theta)$ and $\Delta_\0k=\Delta\phi_\theta$ into the Dyson equation
\begin{align}
g^{-1}-G^{-1}~=~
{\hbox{
\begin{tikzpicture}
    \begin{feynman}
    \vertex (i) ;
    \vertex[right=0.5cm of i] (a);
    \vertex[right=1cm of a] (b);
    \vertex[right=0.5cm of b] (f);
    \vertex[right=0.5cm of a] (c);
    \node[above=0.5cm of c, crossed dot] (d);
    \diagram*{
        (a) -- [fermion] (b),
        (a) -- [scalar] (d) -- [scalar] (b),
    };
    \end{feynman}
\end{tikzpicture}}}
~+~
{\hbox{
\begin{tikzpicture}
    \begin{feynman}
    \vertex (i) ;
    \vertex[right=0.5cm of i] (a);
    \vertex[right=1cm of a] (b);
    \vertex[right=0.5cm of b] (f);
    \vertex[right=0.5cm of a] (c);
    \diagram*{
        (a) -- [fermion] (b) ,
        (a) -- [photon, half left, looseness=2] (b),
    };
    \end{feynman}
\end{tikzpicture}}}
~,
\end{align}
where the dashed and waved lines represent the disorder potential $V^{\rm imp}$ and pairing interaction $V^{\rm pair}$, respectively, and the crossed dot represents the impurity density $n_i$.
After some algebra, we find the single-particle dispersion is unchanged ($\tilde{\varepsilon}_\0k=\varepsilon_\0k$) for particle-hole symmetric (near Fermi energy) systems and another two self-consistent equations
\begin{align}
i\w_n-i\tilde{\w}_n&=-\frac{f_\theta^2 n_iV_i^2}{N} \sum_{\0k'} \frac{i\tilde{\w}_n f_{\theta'}^2}{\tilde{\w}_n^2 + {\varepsilon}_{\0k'}^2+\Delta_{\0k'}^2} , \\
1 &= \frac{VT}{N}\sum_{|\w_n|<\W}\sum_{\0k} \frac{\phi_\theta^2}{\tilde{\w}_n^{2}+{\varepsilon}_{\0k}^2+\Delta_{\0k}^2} ,
\end{align}
where $N$ is the number of $\0k$ (or number of lattice sites).
The sum over $\0k$ can be partially performed through integration in energy (in the wide band approximation), yielding
\begin{align}
i\w_n-i\tilde{\w}_n &= -i f_\theta^{2}\Gamma_i  \int \frac{\md\theta'}{2\pi} \frac{\tilde{\w}_n f_{\theta'}^{2}}{\sqrt{\tilde{\w}_n^2+\Delta^2\phi_{\theta'}^2}} , \label{eq:self0}\\
1 &= \pi\lambda T\sum_{|\w_n|<\W}\int\frac{\md\theta}{2\pi}\frac{\phi_\theta^2}{\sqrt{\tilde{\w}_n^{2}+\Delta^2\phi_\theta^2}} , \label{eq:self1}
\end{align}
where $\lambda=\+NV$ and is set as $0.3$ in the numerical calculations.
From Eq.~\ref{eq:self0}, it is seen that the self-energy $\Sigma(\w_n,\theta)=i\w_n-i\tilde{\w}_n$ depends on $\theta$ only through the form factor $f_\theta^2$.
Hence, we further define $\Sigma(\w_n,\theta)=\Sigma(\w_n)f_\theta^2$.
It can be checked that the normal state self-energy $\Sigma_n(\w_n,\theta)$ can be recovered by setting $\Delta=0$ in Eq.~\ref{eq:self0}.

\begin{figure}
\includegraphics[width=\linewidth]{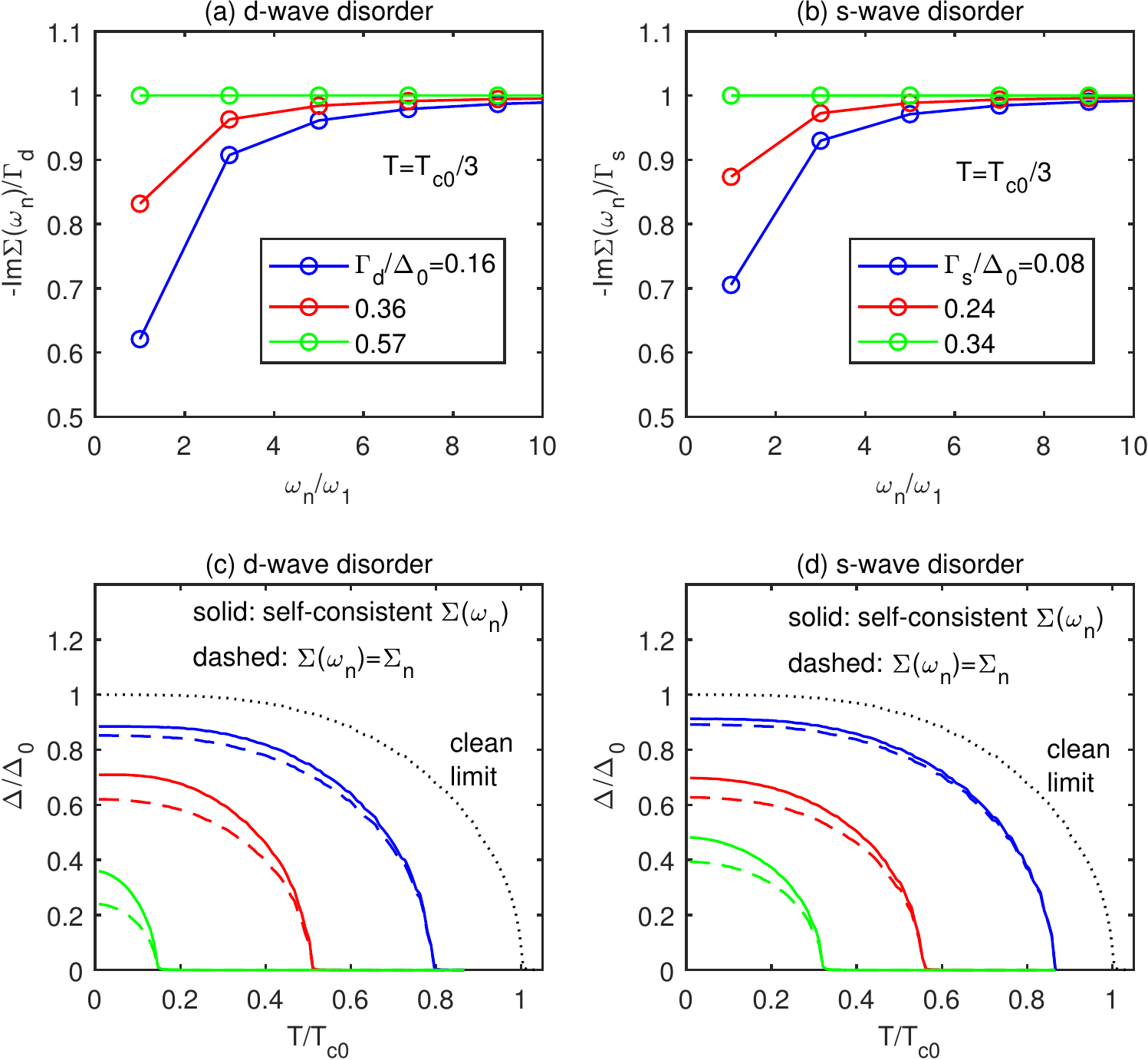}
\caption{
(a) and (b) plot the self-energy $\Sigma(\w_n)$ versus the Matsubara frequency $\w_n$, at a fixed temperature $T=T_{c0}/3$, for d-wave and s-wave disorders, respectively.
(c) and (d) plot the pairing gap $\Delta$ versus the temperature $T$.
The dashed lines are results of the scattering rate approximation with $\Sigma(\w_n)=\Sigma_n$.
The dotted lines are obtained in the clean limit.
}
\label{fig:plot1}
\end{figure}

In the superconducting state, Eq.~\ref{eq:self0} and Eq.~\ref{eq:self1} are combined together to solve $\Sigma(\w_n)$ and $\Delta$ self-consistently.
For the d-wave disorder, the typical results of $\Sigma(\w_n)$, which is purely imaginary, are shown in Fig.~\ref{fig:plot1}(a) at a fixed temperature $T=T_{c0}/3$ ($T_{c0}$ is the transition temperature without disorder).
It can be seen that the amplitude of $\Sigma(\w_n)$ is reduced at low frequencies since low energy quasiparticles are largely gapped out by the d-wave pairing.
This frequency dependence is reduced with increasing $\Gamma_d$ and absent in the dirty limit.
In Fig.~\ref{fig:plot1}(c), we plot the temperature dependence of $\Delta$ (normalized by its zero temperature value without disorder $\Delta_0$).
As a comparison, we also plot the results within the scattering rate approximation $\Sigma(\w_n)=\Sigma_n$.
Clearly, $T_c$ is not affected by the self-consistency of $\Sigma(\w_n)$ since $\Delta=0$ at $T_c$ such that $\Sigma(\w_n)=\Sigma_n$ exactly.
But as temperature decreases, $\Delta$ grows up and reduces $|\Sigma(\w_n)|$ at low frequencies, hence, weakening the pair-breaking effect and enhancing $\Delta$.
The above results are similar for s-wave point disorder as shown in the right panels of Fig.~\ref{fig:plot1}.

\begin{figure}
\includegraphics[width=0.7\linewidth]{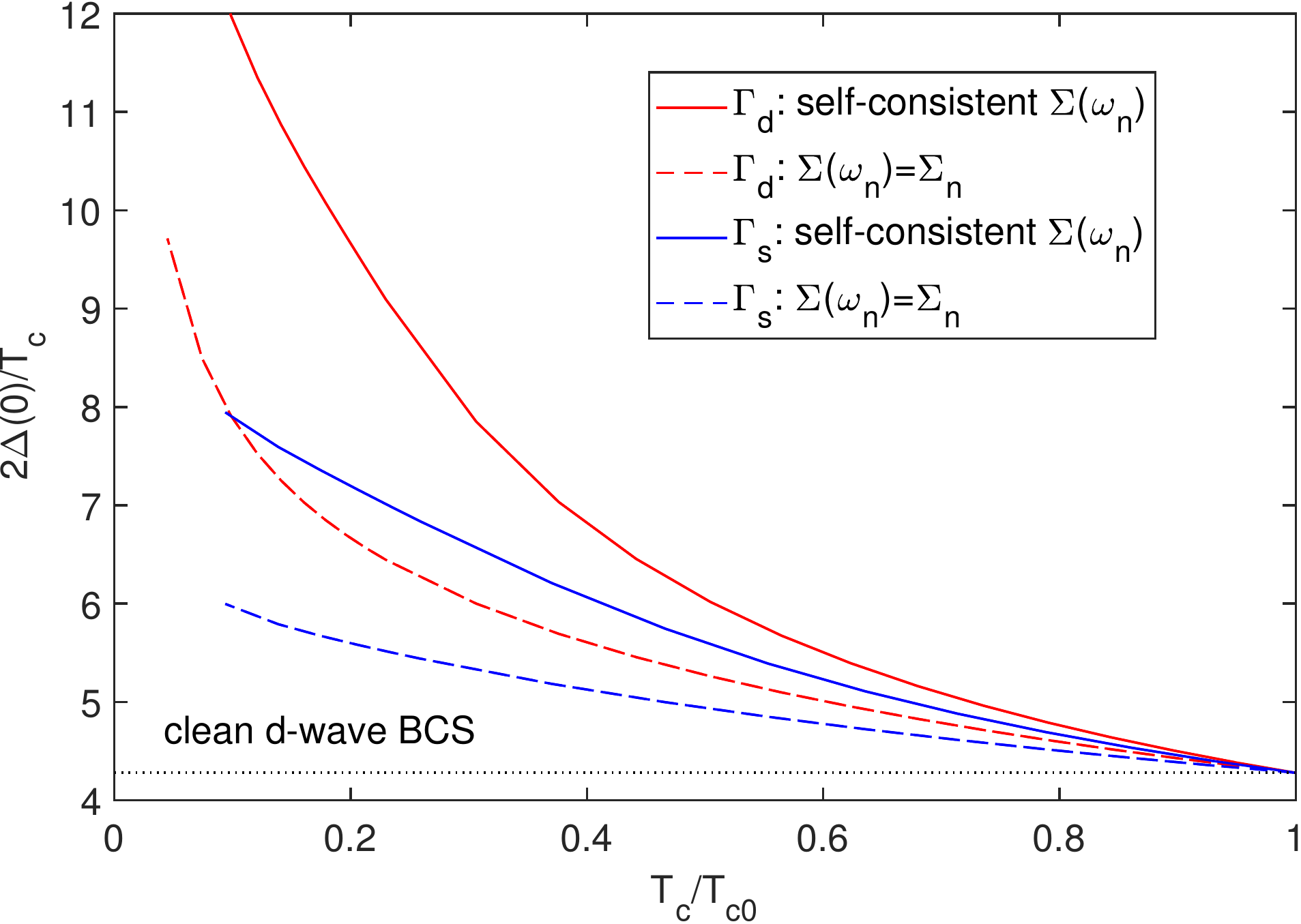}
\caption{
The gap-to-Tc ratio $2\Delta(0)/T_c$ is plotted with respect to $T_c/T_{c0}$.
The red (blue) lines are for d-wave (s-wave) disorders.
The dashed lines are obtained within the scattering rate approximation $\Sigma(\w_n)=\Sigma_n$.
The BCS value $4.28$ in the clean limit is also presented for comparison.
}
\label{fig:ratio}
\end{figure}

One consequence of the disorder effect is to enlarge the gap-to-$T_c$ ratio $2\Delta(0)/T_c$ for both d-wave and s-wave disorders, as shown in Fig.~\ref{fig:ratio}.
The results of scattering rate approximation with $\Sigma(\w_n)=\Sigma_n$ (as we did in Ref.~\cite{Wang2022}) are also given for comparison.
It is clear that the full self-consistent $\Sigma(\w_n)$ further enhances the gap-to-$T_c$ ratio, much larger than the BCS prediction $4.28$ in the clean limit.
This is a strong indication that the large value of this ratio in experiments should not be simply used to identify the strong coupling superconductors with unconventional pairings.

\begin{figure}[b]
\includegraphics[width=\linewidth]{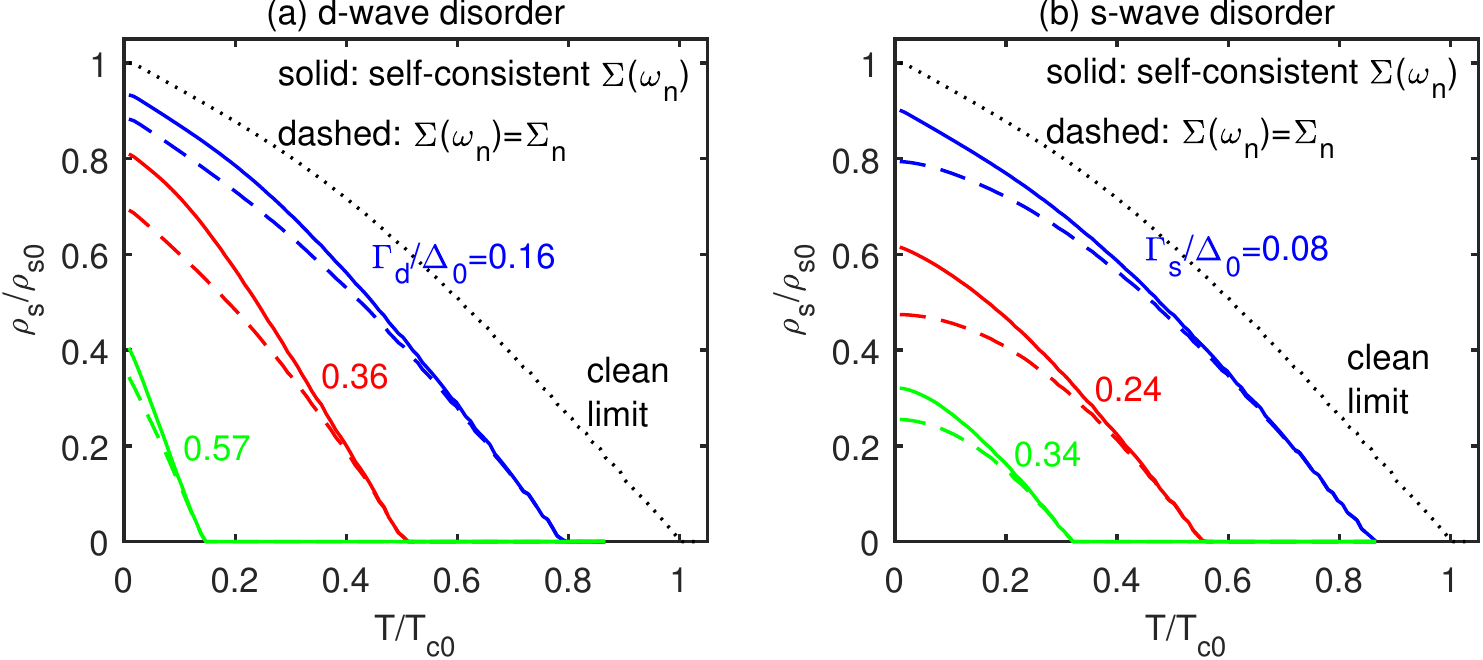}
\caption{
(a) and (b) plot the superfluid density $\rho_s/\rho_{s0}$ versus the temperature $T/T_{c0}$.
The dashed lines are obtained within the scattering rate approximation.
The dotted lines indicate the results without disorder.
}
\label{fig:rhos}
\end{figure}

After $\Sigma(\w_n)$ and $\Delta$ are self-consistently determined, we are in a position to obtain the superfluid density $\rho_s$ as \cite{ColemanBook,Hirschfeld1993}
\begin{align}
\rho_s=2\pi e^2T\+Nv_F^2\sum_{|\w_n|<\Omega}\int\frac{\md\theta}{2\pi} \frac{\cos^2\theta\Delta^2\phi_\theta^2}{[\tilde{\w}_n^2+\Delta^2\phi_\theta^2]^{3/2}} ,
\end{align}
where $e$ is the electron charge and $v_F$ is the Fermi velocity.
The results of $\rho_s$ versus $T$ are shown in Fig.~\ref{fig:rhos}(a) and (b) for d-wave and s-wave disorders, respectively.
For the s-wave disorder, if we simply use $\Sigma(\w_n)=\Sigma_n$, $\rho_s$ always depends on $T$ quadratically at low temperature.
If we use the fully self-consistent $\Sigma(\w_n)$, the $T$-dependence remains to be almost linear for small $\Gamma_s$ near the clean limit, but still exhibits quadratic power law as approaching the dirty limit \cite{Lee-Hone2017}, hence, inconsistent with the experiment \cite{Bozovic2016}.
On the other hand, for the d-wave disorder, $\rho_s$ always depends on $T$ linearly at low temperature, whether the frequency dependence of $\Sigma(\w_n)$ is considered or not, confirming our previous conclusion under the scattering rate approximation \cite{Wang2022}.

\begin{figure}
\includegraphics[width=\linewidth]{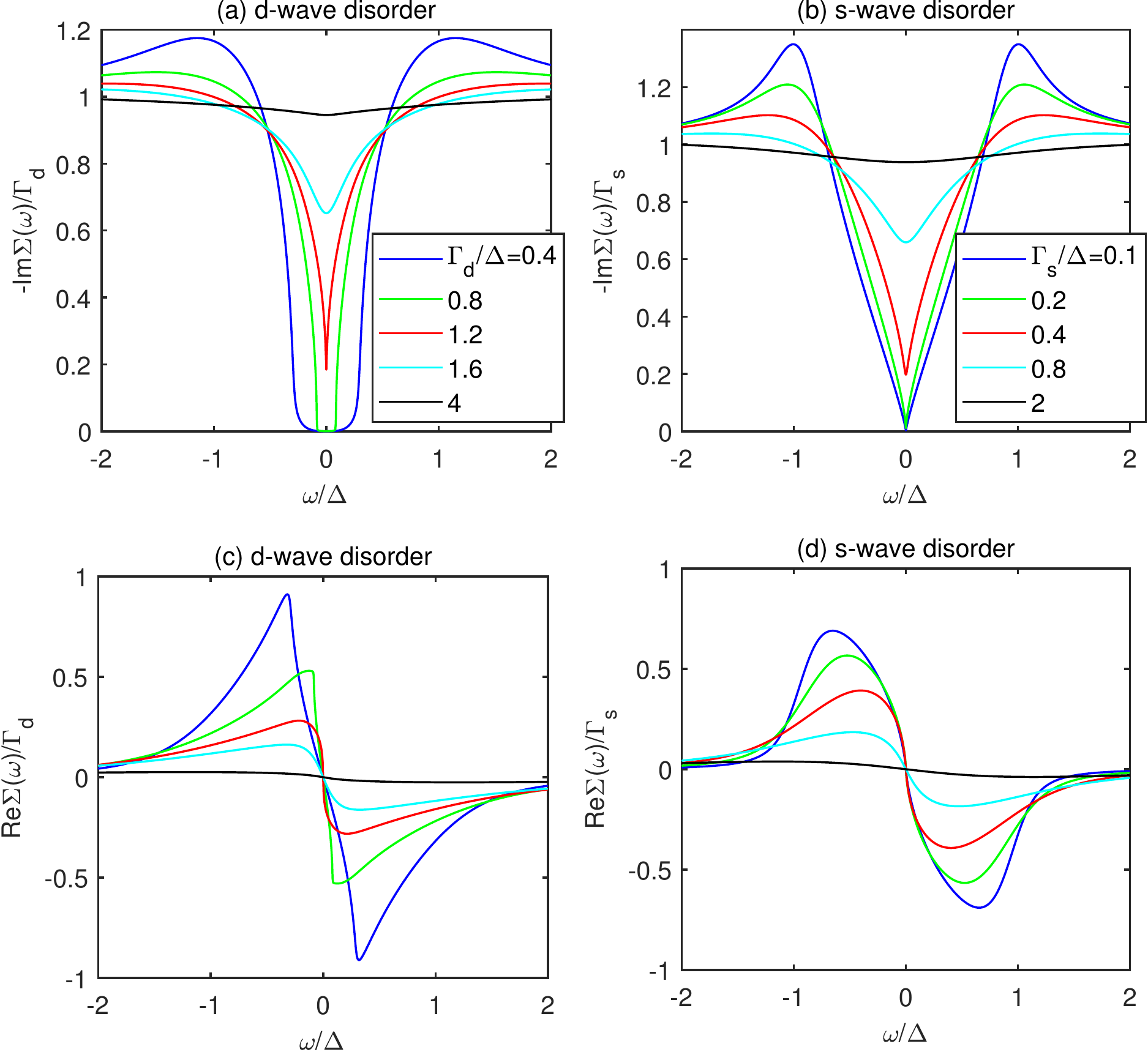}
\caption{
(a) and (b) plot the imaginary part of the self-energy $-\mathrm{Im}\Sigma(\w)$ versus frequency $\w$ for d-wave and s-wave disorders, respectively.
(c) and (d) are results of the real part  $\mathrm{Re}\Sigma(\w)$.
}
\label{fig:selfEn}
\end{figure}

In order to achieve better understanding of the d-wave disorder, we move to study real frequency quantities.
Since $\Delta$ is frequency independent within the BCS approximation, we only need to further solve the retarded self-energy $\Sigma(\w,\theta)$, which can be obtained by analytic continuation of Eq.~\ref{eq:self0}, leading to
\begin{align}
\tilde{\w}=\w+i0^+ + f_{\theta}^{2}\Gamma_i \int \frac{\md\theta'}{2\pi} \frac{\tilde{\w} f_{\theta'}^{2}}{\sqrt{-\tilde{\w}^2+\Delta^2\phi_{\theta'}^2}} , \label{eq:self0Real}
\end{align}
where $\tilde{\w}(\w,\theta)=\w-\Sigma(\w,\theta)$ with $\Sigma(\w,\theta)=\Sigma(\w)f_\theta^2$.
In the real frequency domain, the normal state self-energy becomes $\Sigma_n(\w,\theta)=-i\Gamma_if_\theta^2=\Sigma_n(\w)f_\theta^2$.
In the superconducting state, Eq.~\ref{eq:self0Real} can be used to solve $\Sigma(\w)$ self-consistently for each given $\w$ and $\Delta$.
In Fig.~\ref{fig:selfEn}, we plot both the real and imaginary parts of $\Sigma(\w)$ in (a,b) and (c,d), respectively.
Let us first look at the s-wave disorder.
As shown in Fig.~\ref{fig:selfEn}(b), $-\mathrm{Im}\Sigma$ drops to small values almost linearly with $|\w|$ for small $\Gamma_s$, exhibiting a V-shape behavior.
As $\Gamma_s$ increases, $-\mathrm{Im}\Sigma$ tends to be the constant $\Gamma_s$ as anticipated in the dirty limit.
Meanwhile, the real part of $\Sigma(\w)$, shown in Fig.~\ref{fig:selfEn}(d), shows the corresponding frequency dependence for $\w<\Delta$ as revealed by the Kronig-Kramers relations, and tends to vanish in the dirty limit.
These behaviors are similar for the d-wave disorder as shown in Fig.~\ref{fig:selfEn}(a,c), except that for small $\Gamma_d$, $-\mathrm{Im}\Sigma$ shows a U-shape dependence on $\w$ and its coherence peak at $\w\sim\Delta$ becomes much smoother as shown in Fig.~\ref{fig:selfEn}(a).
The vanishment of $-\mathrm{Im}\Sigma(\w)$ indicates the low energy quasiparticles feel almost no scattering rate.

\begin{figure}
\includegraphics[width=\linewidth]{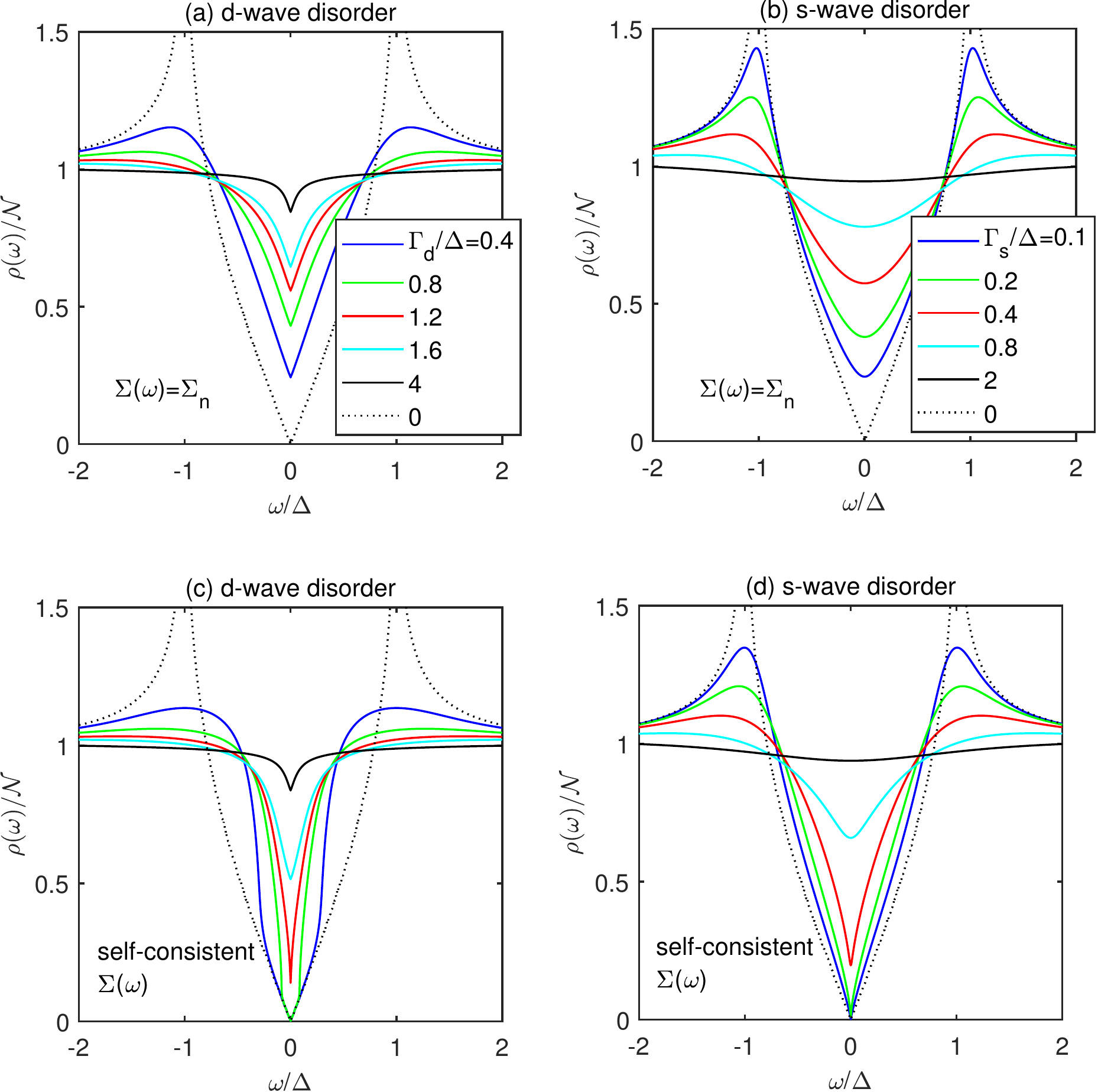}
\caption{
(a) and (b) plot the DOS $\rho(\w)$ within the scattering rate approximation for the d-wave and s-wave disorders, respectively.
(c) and (d) are similar to (a) and (b) but obtained with the self-consistent $\Sigma(\w)$.
}
\label{fig:dos}
\end{figure}

After $\Sigma(\w)$ is obtained, the imaginary part of the retarded Green's function $G_{11}(\w)$ gives the DOS $\rho(\w)$ as
\begin{align}
\rho(\w)=\+N\int\frac{\md\theta}{2\pi}\text{Im}\frac{\tilde{\w}}{\sqrt{-\tilde{\w}^2+\Delta^2\phi_\theta^2}} , \label{eq:dos}
\end{align}
where $\tilde{\w}=\w-\Sigma(\w)f_\theta^2$.
The results of $\rho(\w)$ are plotted in Fig.~\ref{fig:dos} with the scattering rate approximation in (a,b) and with the self-consistent $\Sigma(\w)$ in (c,d).
For the s-wave disorder, any nonzero $\Gamma_s$ under the scattering approximation causes $\rho(\w)-\rho(0)\propto\w^2$ as shown in Fig.~\ref{fig:dos}(b).
After self-consistency of $\Sigma(\w)$, the $\w$-dependence persists to be almost linear for small $\Gamma_s$, and becomes quadratic for large $\Gamma_s$ as shown in Fig.~\ref{fig:dos}(d).
This explains the temperature dependence of $\rho_s(T)$ in Fig.~\ref{fig:rhos}(b) as discussed above since dropping of $\rho_s$ with $T$ is contributed by quasiparticle excitations with energy $\w\sim T$.
Then let us look at the d-wave disorder.
With $\Sigma(\w)=\Sigma_n$, $\rho(\w)-\rho(0)\propto|\w|$ at small $|\w|$ for all $\Gamma_d$ as shown in Fig.~\ref{fig:dos}(a).
This linear dependence does not change after self-consistency as shown in Fig.~\ref{fig:dos}(c), also consistent with the superfluid density $\rho_s(T)$ in Fig.~\ref{fig:rhos}(a).
Furthermore, it is interesting to find that for small $\Gamma_d$, the DOS $\rho(\w)$ falls onto the clean limit curve at low energy which is consistent with the vanishement of $-\mathrm{Im}\Sigma(\w)$ as shown in Fig.~\ref{fig:selfEn}(a).

\begin{figure}
\includegraphics[width=0.7\linewidth]{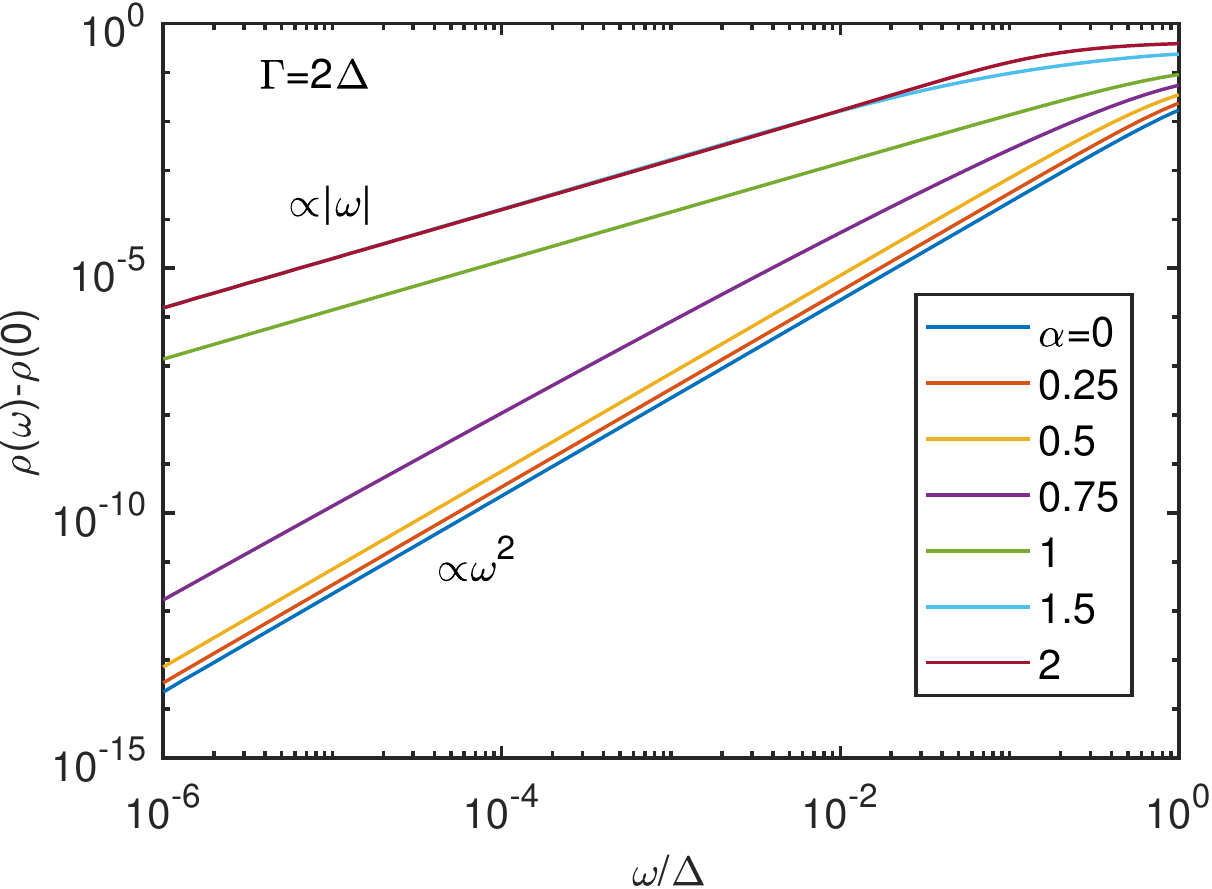}
\caption{ Universal scalings of the DOS $\rho(\w)$ for the ``soft'' scattering rate defined in Eq.~\ref{eq:softDisorder}.
}
\label{fig:dosScaling}
\end{figure}

From the above, we have shown that the self-consistency of $\Sigma(\w)$ can change the low energy scaling behavior of $\rho(\w)-\rho(0)$ from $\w^2$ in the dirty limit to $|\w|$ in the clean limit for the s-wave disorder, while does not change the scaling (always $|\w|$) for the d-wave disorder.
The essential reason is that it drops to zero along each nodal direction $\theta_0$ and hence has little effect on low energy quasiparticles.
In the following, to achieve more universal results, we consider a generalized ``soft'' scattering rate near each nodal direction $\theta_0$, given by
\begin{align}\label{eq:softDisorder}
\Sigma(\w,\theta)=-i\Gamma_\alpha|\theta-\theta_0|^\alpha .
\end{align}
In fact, any scattering rate can be expanded near each $\theta_0$ and thus be described by the above model.
For the s-wave and d-wave disorders discussed above, $\alpha=0$ and $2$, respectively.
At low energy, we only need to consider $\theta$ near each $\theta_0$ such that $\phi_\theta=\cos(2\theta)\sim\pm2(\theta-\theta_0)$.
By further shifting $\theta-\theta_0$ to $\theta$, Eq.~\ref{eq:dos} becomes
\begin{align}
\rho(\w)\sim\+N\int_0^{\Theta}\frac{\md\theta}{\Theta}\text{Im}\frac{\w+i\Gamma_\alpha\theta^\alpha}{\sqrt{-(\w+i\Gamma_\alpha\theta^\alpha)^2+4\Delta^2\theta^2}} ,
\end{align}
where $\Theta$ is an angle cutoff which does not change the low energy scaling behavior.
In the limit of $\w\to0$, we complete the angle integral numerically to find that there are only two universal scalings
\begin{align}
\rho(\w)-\rho(0)\propto\left\{ \begin{array}{ll}
\w^2, & \alpha<1 , \\
|\w|, & \alpha\ge1 ,
\end{array}  \right.
\end{align}
as shown in Fig.~\ref{fig:dosScaling}.
Clearly, both the results of s-wave disorder ($\alpha=0$) and d-wave disorder ($\alpha=2$) are correctly captured.
Based on the DOS, the scalings of some other physical quantities are expected.
For example, the entropy, specific heat $C/T$, superfluid density $\rho_s$ and penetration depth all should depend on $T$ quadratically for $\alpha<1$ and linearly for $\alpha\ge1$ in the low temperature limit.

In summary, we have studied the self-energy effect of the d-wave disorder in d-wave superconductors, mainly focusing on comparison with the scattering rate approximation which tends to become exact in the dirty limit.
The gap-to-$T_c$ ratio is found to be much larger than the BCS value in the clean limit.
The DOS $\rho(\w)-\rho(0)$ at low energy is found to exhibit a linear scaling behavior $\propto|\w|$ for all $\Gamma_d$ and a quadratic scaling $\propto\w^2$ for not too small $\Gamma_s$.
Within the scattering rate approximation, these two scalings are then generalized to a more general ``soft'' scattering rate $\Gamma_\theta= \Gamma_\alpha |\theta-\theta_0|^\alpha$ which gives $\rho(\w)-\rho(0)\propto|\w|$ ($\w^2$) for $\alpha\ge1$ ($\alpha<1$) falling into two and only two categories.

Finally, we make some remarks on experiments.
(1) In the dirty limit, the low energy scaling behavior of the DOS $\rho(\w)$ is quite different for the d-wave (linearly) and s-wave (quadratically) disorders, hence, leading to the fundamental difference for the temperature dependence of some thermodynamic quantities such as the entropy, specific heat $C/T$, superfluid density $\rho_s$ and penetration depth.
(2) In the clean limit, both types of disorders give the same (almost linear) scaling at low energy, but the coherence peak at high energy is largely smoothed for the d-wave disorder, which is different from the s-wave disorder.
(3) In real materials, the two types of disorders can coexist. For the low energy region $(T,\omega)\ll\Gamma_s$, the s-wave disorder effect dominates. But for the intermediate region with $\Gamma_s\ll (T,\omega)\ll\Gamma_d$, the d-wave disorder effect dominates.

This work is supported by National Key R\&D Program of China (Grant No. 2022YFA1403201) and National Natural Science Foundation of China (Grant No. 12274205 and No. 11874205).

\bibliography{disorder}

\begin{thebibliography}{19}%
\makeatletter
\providecommand \@ifxundefined [1]{%
 \@ifx{#1\undefined}
}%
\providecommand \@ifnum [1]{%
 \ifnum #1\expandafter \@firstoftwo
 \else \expandafter \@secondoftwo
 \fi
}%
\providecommand \@ifx [1]{%
 \ifx #1\expandafter \@firstoftwo
 \else \expandafter \@secondoftwo
 \fi
}%
\providecommand \natexlab [1]{#1}%
\providecommand \enquote  [1]{``#1''}%
\providecommand \bibnamefont  [1]{#1}%
\providecommand \bibfnamefont [1]{#1}%
\providecommand \citenamefont [1]{#1}%
\providecommand \href@noop [0]{\@secondoftwo}%
\providecommand \href [0]{\begingroup \@sanitize@url \@href}%
\providecommand \@href[1]{\@@startlink{#1}\@@href}%
\providecommand \@@href[1]{\endgroup#1\@@endlink}%
\providecommand \@sanitize@url [0]{\catcode `\\12\catcode `\$12\catcode
  `\&12\catcode `\#12\catcode `\^12\catcode `\_12\catcode `\%12\relax}%
\providecommand \@@startlink[1]{}%
\providecommand \@@endlink[0]{}%
\providecommand \url  [0]{\begingroup\@sanitize@url \@url }%
\providecommand \@url [1]{\endgroup\@href {#1}{\urlprefix }}%
\providecommand \urlprefix  [0]{URL }%
\providecommand \Eprint [0]{\href }%
\providecommand \doibase [0]{https://doi.org/}%
\providecommand \selectlanguage [0]{\@gobble}%
\providecommand \bibinfo  [0]{\@secondoftwo}%
\providecommand \bibfield  [0]{\@secondoftwo}%
\providecommand \translation [1]{[#1]}%
\providecommand \BibitemOpen [0]{}%
\providecommand \bibitemStop [0]{}%
\providecommand \bibitemNoStop [0]{.\EOS\space}%
\providecommand \EOS [0]{\spacefactor3000\relax}%
\providecommand \BibitemShut  [1]{\csname bibitem#1\endcsname}%
\let\auto@bib@innerbib\@empty
\bibitem [{\citenamefont {Anderson}(1959)}]{Anderson1959}%
  \BibitemOpen
  \bibfield  {author} {\bibinfo {author} {\bibfnamefont {P.~W.}\ \bibnamefont
  {Anderson}},\ }\bibfield  {title} {\bibinfo {title} {Theory of dirty
  superconductors},\ }\href {https://doi.org/10.1016/0022-3697(59)90036-8}
  {\bibfield  {journal} {\bibinfo  {journal} {Journal of Physics and Chemistry
  of Solids}\ }\textbf {\bibinfo {volume} {11}},\ \bibinfo {pages} {26}
  (\bibinfo {year} {1959})}\BibitemShut {NoStop}%
\bibitem [{\citenamefont {Abrikosov}\ and\ \citenamefont
  {Gor'kov}(1958)}]{AG1958}%
  \BibitemOpen
  \bibfield  {author} {\bibinfo {author} {\bibfnamefont {A.~A.}\ \bibnamefont
  {Abrikosov}}\ and\ \bibinfo {author} {\bibfnamefont {L.~P.}\ \bibnamefont
  {Gor'kov}},\ }\bibfield  {title} {\bibinfo {title} {On the theory of
  superconducting alloys},\ }\href@noop {} {\bibfield  {journal} {\bibinfo
  {journal} {Sov. Phys. JETP}\ }\textbf {\bibinfo {volume} {35}},\ \bibinfo
  {pages} {1090} (\bibinfo {year} {1958})}\BibitemShut {NoStop}%
\bibitem [{\citenamefont {Abrikosov}\ and\ \citenamefont
  {Gor'kov}(1959)}]{AG1959}%
  \BibitemOpen
  \bibfield  {author} {\bibinfo {author} {\bibfnamefont {A.~A.}\ \bibnamefont
  {Abrikosov}}\ and\ \bibinfo {author} {\bibfnamefont {L.~P.}\ \bibnamefont
  {Gor'kov}},\ }\bibfield  {title} {\bibinfo {title} {Superconducting alloys at
  finite temperatures},\ }\href@noop {} {\bibfield  {journal} {\bibinfo
  {journal} {Sov. Phys. JETP}\ }\textbf {\bibinfo {volume} {36}},\ \bibinfo
  {pages} {319} (\bibinfo {year} {1959})}\BibitemShut {NoStop}%
\bibitem [{\citenamefont {Yu}(1965)}]{Yu1965}%
  \BibitemOpen
  \bibfield  {author} {\bibinfo {author} {\bibfnamefont {L.}~\bibnamefont
  {Yu}},\ }\bibfield  {title} {\bibinfo {title} {superconductors with magnetic
  impurities},\ }\href@noop {} {\bibfield  {journal} {\bibinfo  {journal} {Acta
  Phys. Sin.}\ }\textbf {\bibinfo {volume} {21}},\ \bibinfo {pages} {75}
  (\bibinfo {year} {1965})}\BibitemShut {NoStop}%
\bibitem [{\citenamefont {Abrikosov}\ and\ \citenamefont
  {Gor'kov}(1961)}]{AG1961}%
  \BibitemOpen
  \bibfield  {author} {\bibinfo {author} {\bibfnamefont {A.~A.}\ \bibnamefont
  {Abrikosov}}\ and\ \bibinfo {author} {\bibfnamefont {L.~P.}\ \bibnamefont
  {Gor'kov}},\ }\bibfield  {title} {\bibinfo {title} {Contribution to the
  theory of superconducting alloys with paramagnetic impurities},\ }\href@noop
  {} {\bibfield  {journal} {\bibinfo  {journal} {Sov. Phys. JETP}\ }\textbf
  {\bibinfo {volume} {12}},\ \bibinfo {pages} {1243} (\bibinfo {year}
  {1961})}\BibitemShut {NoStop}%
\bibitem [{\citenamefont {Alloul}\ \emph {et~al.}(2009)\citenamefont {Alloul},
  \citenamefont {Bobroff}, \citenamefont {Gabay},\ and\ \citenamefont
  {Hirschfeld}}]{RMP2009}%
  \BibitemOpen
  \bibfield  {author} {\bibinfo {author} {\bibfnamefont {H.}~\bibnamefont
  {Alloul}}, \bibinfo {author} {\bibfnamefont {J.}~\bibnamefont {Bobroff}},
  \bibinfo {author} {\bibfnamefont {M.}~\bibnamefont {Gabay}},\ and\ \bibinfo
  {author} {\bibfnamefont {P.}~\bibnamefont {Hirschfeld}},\ }\bibfield  {title}
  {\bibinfo {title} {Defects in correlated metals and superconductors},\ }\href
  {https://doi.org/10.1103/RevModPhys.81.45} {\bibfield  {journal} {\bibinfo
  {journal} {Rev. Mod. Phys.}\ }\textbf {\bibinfo {volume} {81}},\ \bibinfo
  {pages} {45} (\bibinfo {year} {2009})}\BibitemShut {NoStop}%
\bibitem [{\citenamefont {Xiang}\ and\ \citenamefont {Wu}(2022)}]{XiangBook}%
  \BibitemOpen
  \bibfield  {author} {\bibinfo {author} {\bibfnamefont {T.}~\bibnamefont
  {Xiang}}\ and\ \bibinfo {author} {\bibfnamefont {C.}~\bibnamefont {Wu}},\
  }\href {https://doi.org/10.1017/9781009218566} {\emph {\bibinfo {title}
  {D-wave Superconductivity}}}\ (\bibinfo  {publisher} {Cambridge University
  Press},\ \bibinfo {year} {2022})\BibitemShut {NoStop}%
\bibitem [{\citenamefont {Božović}\ \emph {et~al.}(2016)\citenamefont
  {Božović}, \citenamefont {He}, \citenamefont {Wu},\ and\ \citenamefont
  {Bollinger}}]{Bozovic2016}%
  \BibitemOpen
  \bibfield  {author} {\bibinfo {author} {\bibfnamefont {I.}~\bibnamefont
  {Božović}}, \bibinfo {author} {\bibfnamefont {X.}~\bibnamefont {He}},
  \bibinfo {author} {\bibfnamefont {J.}~\bibnamefont {Wu}},\ and\ \bibinfo
  {author} {\bibfnamefont {A.~T.}\ \bibnamefont {Bollinger}},\ }\bibfield
  {title} {\bibinfo {title} {Dependence of the critical temperature in
  overdoped copper oxides on superfluid density},\ }\href
  {https://doi.org/10.1038/nature19061} {\bibfield  {journal} {\bibinfo
  {journal} {Nature}\ }\textbf {\bibinfo {volume} {536}},\ \bibinfo {pages}
  {309} (\bibinfo {year} {2016})}\BibitemShut {NoStop}%
\bibitem [{\citenamefont {Mahmood}\ \emph {et~al.}(2019)\citenamefont
  {Mahmood}, \citenamefont {He}, \citenamefont {Božović},\ and\ \citenamefont
  {Armitage}}]{Mahmood2019}%
  \BibitemOpen
  \bibfield  {author} {\bibinfo {author} {\bibfnamefont {F.}~\bibnamefont
  {Mahmood}}, \bibinfo {author} {\bibfnamefont {X.}~\bibnamefont {He}},
  \bibinfo {author} {\bibfnamefont {I.}~\bibnamefont {Božović}},\ and\
  \bibinfo {author} {\bibfnamefont {N.~P.}\ \bibnamefont {Armitage}},\
  }\bibfield  {title} {\bibinfo {title} {Locating the missing superconducting
  electrons in the overdoped cuprates {La$_{2-x}$Sr$_x$CuO$_4$}},\ }\href
  {https://doi.org/10.1103/PhysRevLett.122.027003} {\bibfield  {journal}
  {\bibinfo  {journal} {Phys. Rev. Lett.}\ }\textbf {\bibinfo {volume} {122}},\
  \bibinfo {pages} {027003} (\bibinfo {year} {2019})}\BibitemShut {NoStop}%
\bibitem [{\citenamefont {Wang}\ \emph {et~al.}(2022)\citenamefont {Wang},
  \citenamefont {Xu}, \citenamefont {Zhang},\ and\ \citenamefont
  {Wang}}]{Wang2022}%
  \BibitemOpen
  \bibfield  {author} {\bibinfo {author} {\bibfnamefont {D.}~\bibnamefont
  {Wang}}, \bibinfo {author} {\bibfnamefont {J.-Q.}\ \bibnamefont {Xu}},
  \bibinfo {author} {\bibfnamefont {H.-J.}\ \bibnamefont {Zhang}},\ and\
  \bibinfo {author} {\bibfnamefont {Q.-H.}\ \bibnamefont {Wang}},\ }\bibfield
  {title} {\bibinfo {title} {Anisotropic {{Scattering Caused}} by {{Apical
  Oxygen Vacancies}} in {{Thin Films}} of {{Overdoped High-Temperature Cuprate
  Superconductors}}},\ }\href {https://doi.org/10.1103/PhysRevLett.128.137001}
  {\bibfield  {journal} {\bibinfo  {journal} {Phys. Rev. Lett.}\ }\textbf
  {\bibinfo {volume} {128}},\ \bibinfo {pages} {137001} (\bibinfo {year}
  {2022})}\BibitemShut {NoStop}%
\bibitem [{\citenamefont {Ioffe}\ and\ \citenamefont
  {Millis}(1998)}]{Ioffe1998}%
  \BibitemOpen
  \bibfield  {author} {\bibinfo {author} {\bibfnamefont {L.~B.}\ \bibnamefont
  {Ioffe}}\ and\ \bibinfo {author} {\bibfnamefont {A.~J.}\ \bibnamefont
  {Millis}},\ }\bibfield  {title} {\bibinfo {title} {Zone-diagonal-dominated
  transport in high-{$T_c$} cuprates},\ }\href
  {https://doi.org/10.1103/PhysRevB.58.11631} {\bibfield  {journal} {\bibinfo
  {journal} {Phys. Rev. B}\ }\textbf {\bibinfo {volume} {58}},\ \bibinfo
  {pages} {11631} (\bibinfo {year} {1998})}\BibitemShut {NoStop}%
\bibitem [{\citenamefont {Sato}\ \emph {et~al.}(2000)\citenamefont {Sato},
  \citenamefont {Tsukada}, \citenamefont {Naito},\ and\ \citenamefont
  {Matsuda}}]{Sato2000}%
  \BibitemOpen
  \bibfield  {author} {\bibinfo {author} {\bibfnamefont {H.}~\bibnamefont
  {Sato}}, \bibinfo {author} {\bibfnamefont {A.}~\bibnamefont {Tsukada}},
  \bibinfo {author} {\bibfnamefont {M.}~\bibnamefont {Naito}},\ and\ \bibinfo
  {author} {\bibfnamefont {A.}~\bibnamefont {Matsuda}},\ }\bibfield  {title}
  {\bibinfo {title}
  {{${\mathrm{La}}_{2\ensuremath{-}x}{\mathrm{Sr}}_{x}{\mathrm{CuO}}_{y}$}
  epitaxial thin films ($x=0$ to $2$): {Structure}, strain, and
  superconductivity},\ }\href {https://doi.org/10.1103/PhysRevB.61.12447}
  {\bibfield  {journal} {\bibinfo  {journal} {Phys. Rev. B}\ }\textbf {\bibinfo
  {volume} {61}},\ \bibinfo {pages} {12447} (\bibinfo {year}
  {2000})}\BibitemShut {NoStop}%
\bibitem [{\citenamefont {Kim}\ \emph {et~al.}(2017)\citenamefont {Kim},
  \citenamefont {Christiani}, \citenamefont {Logvenov}, \citenamefont {Choi},
  \citenamefont {Kim}, \citenamefont {Minola},\ and\ \citenamefont
  {Keimer}}]{Kim2017}%
  \BibitemOpen
  \bibfield  {author} {\bibinfo {author} {\bibfnamefont {G.}~\bibnamefont
  {Kim}}, \bibinfo {author} {\bibfnamefont {G.}~\bibnamefont {Christiani}},
  \bibinfo {author} {\bibfnamefont {G.}~\bibnamefont {Logvenov}}, \bibinfo
  {author} {\bibfnamefont {S.}~\bibnamefont {Choi}}, \bibinfo {author}
  {\bibfnamefont {H.-H.}\ \bibnamefont {Kim}}, \bibinfo {author} {\bibfnamefont
  {M.}~\bibnamefont {Minola}},\ and\ \bibinfo {author} {\bibfnamefont
  {B.}~\bibnamefont {Keimer}},\ }\bibfield  {title} {\bibinfo {title}
  {Selective formation of apical oxygen vacancies in
  {La$_{2-x}$Sr$_x$CuO$_4$}},\ }\href
  {https://doi.org/10.1103/PhysRevMaterials.1.054801} {\bibfield  {journal}
  {\bibinfo  {journal} {Phys. Rev. Materials}\ }\textbf {\bibinfo {volume}
  {1}},\ \bibinfo {pages} {054801} (\bibinfo {year} {2017})}\BibitemShut
  {NoStop}%
\bibitem [{\citenamefont {Abrikosov}\ \emph {et~al.}(1963)\citenamefont
  {Abrikosov}, \citenamefont {Gorkov},\ and\ \citenamefont
  {Dzialoshinskii}}]{AGDBook}%
  \BibitemOpen
  \bibfield  {author} {\bibinfo {author} {\bibfnamefont {A.~A.}\ \bibnamefont
  {Abrikosov}}, \bibinfo {author} {\bibfnamefont {L.}~\bibnamefont {Gorkov}},\
  and\ \bibinfo {author} {\bibfnamefont {I.~E.}\ \bibnamefont
  {Dzialoshinskii}},\ }\href@noop {} {\emph {\bibinfo {title} {Methods of
  quantum field theory in statistical physics}}}\ (\bibinfo  {publisher} {NJ,
  Prentice-Hall},\ \bibinfo {year} {1963})\BibitemShut {NoStop}%
\bibitem [{\citenamefont {Eliashberg}(1960)}]{Eliashberg1960}%
  \BibitemOpen
  \bibfield  {author} {\bibinfo {author} {\bibfnamefont {G.}~\bibnamefont
  {Eliashberg}},\ }\bibfield  {title} {\bibinfo {title} {Interactions between
  electrons and lattice vibrations in a superconductor},\ }\href@noop {}
  {\bibfield  {journal} {\bibinfo  {journal} {Sov. Phys. JETP}\ }\textbf
  {\bibinfo {volume} {11}},\ \bibinfo {pages} {696} (\bibinfo {year}
  {1960})}\BibitemShut {NoStop}%
\bibitem [{\citenamefont {Bardeen}\ \emph {et~al.}(1957)\citenamefont
  {Bardeen}, \citenamefont {Cooper},\ and\ \citenamefont
  {Schrieffer}}]{BCS1957}%
  \BibitemOpen
  \bibfield  {author} {\bibinfo {author} {\bibfnamefont {J.}~\bibnamefont
  {Bardeen}}, \bibinfo {author} {\bibfnamefont {L.~N.}\ \bibnamefont
  {Cooper}},\ and\ \bibinfo {author} {\bibfnamefont {J.~R.}\ \bibnamefont
  {Schrieffer}},\ }\bibfield  {title} {\bibinfo {title} {Theory of
  {{Superconductivity}}},\ }\href {https://doi.org/10.1103/PhysRev.108.1175}
  {\bibfield  {journal} {\bibinfo  {journal} {Phys. Rev.}\ }\textbf {\bibinfo
  {volume} {108}},\ \bibinfo {pages} {1175} (\bibinfo {year}
  {1957})}\BibitemShut {NoStop}%
\bibitem [{\citenamefont {Coleman}(2015)}]{ColemanBook}%
  \BibitemOpen
  \bibfield  {author} {\bibinfo {author} {\bibfnamefont {P.}~\bibnamefont
  {Coleman}},\ }\href {https://doi.org/10.1017/CBO9781139020916} {\emph
  {\bibinfo {title} {Introduction to Many-Body Physics}}}\ (\bibinfo
  {publisher} {Cambridge University Press},\ \bibinfo {year}
  {2015})\BibitemShut {NoStop}%
\bibitem [{\citenamefont {Hirschfeld}\ and\ \citenamefont
  {Goldenfeld}(1993)}]{Hirschfeld1993}%
  \BibitemOpen
  \bibfield  {author} {\bibinfo {author} {\bibfnamefont {P.~J.}\ \bibnamefont
  {Hirschfeld}}\ and\ \bibinfo {author} {\bibfnamefont {N.}~\bibnamefont
  {Goldenfeld}},\ }\bibfield  {title} {\bibinfo {title} {Effect of strong
  scattering on the low-temperature penetration depth of a d-wave
  superconductor},\ }\href {https://doi.org/10.1103/PhysRevB.48.4219}
  {\bibfield  {journal} {\bibinfo  {journal} {Phys. Rev. B}\ }\textbf {\bibinfo
  {volume} {48}},\ \bibinfo {pages} {4219} (\bibinfo {year}
  {1993})}\BibitemShut {NoStop}%
\bibitem [{\citenamefont {Lee-Hone}\ \emph {et~al.}(2017)\citenamefont
  {Lee-Hone}, \citenamefont {Dodge},\ and\ \citenamefont
  {Broun}}]{Lee-Hone2017}%
  \BibitemOpen
  \bibfield  {author} {\bibinfo {author} {\bibfnamefont {N.~R.}\ \bibnamefont
  {Lee-Hone}}, \bibinfo {author} {\bibfnamefont {J.~S.}\ \bibnamefont
  {Dodge}},\ and\ \bibinfo {author} {\bibfnamefont {D.~M.}\ \bibnamefont
  {Broun}},\ }\bibfield  {title} {\bibinfo {title} {Disorder and superfluid
  density in overdoped cuprate superconductors},\ }\href
  {https://doi.org/10.1103/PhysRevB.96.024501} {\bibfield  {journal} {\bibinfo
  {journal} {Phys. Rev. B}\ }\textbf {\bibinfo {volume} {96}},\ \bibinfo
  {pages} {024501} (\bibinfo {year} {2017})}\BibitemShut {NoStop}%
\end{thebibliography}%
\end{document}